\documentclass[aps,pra,english,twocolumn,showpacs,preprintnumbers,floatfix,superscriptaddress]{revtex4-1}
\usepackage[T1]{fontenc}
\usepackage{graphicx}
\usepackage{amsmath,amsthm,amssymb,bm}
\usepackage{verbatim}
\makeatletter

\usepackage{esint}

\usepackage[version=3]{mhchem}

\usepackage{color}

\usepackage{ulem}
\usepackage{hyperref} % this should be loaded after all other packages

\begin{document}

\title{Anisotropic contribution to the van der Waals and the Casimir-Polder energies for CO$_2$ and CH$_4$ molecules near surfaces and thin films}

\author{Priyadarshini Thiyam}
\email{thiyam@kth.se}
\affiliation{Department of Materials Science and Engineering, Royal Institute of Technology, SE-100 44 Stockholm, Sweden}

\author{Prachi Parashar}
\email{prachi@nhn.ou.edu}
\affiliation{Department of Physics, Southern Illinois University-Carbondale, Carbondale, Illinois 62901 USA}
\affiliation{Homer L. Dodge Department of Physics and Astronomy, University of Oklahoma, Norman, Oklahoma 73019, USA}

\author{K. V. Shajesh}
\email{kvshajesh@gmail.com}
\affiliation{Department of Physics, Southern Illinois University-Carbondale, Carbondale, Illinois 62901 USA}

\author{Clas Persson}
\affiliation{Department of Materials Science and Engineering, Royal Institute of Technology, SE-100 44 Stockholm, Sweden}
\affiliation{Department of Physics, University of Oslo, P. O. Box 1048 Blindern, NO-0316 Oslo, Norway}
\affiliation{Centre for Materials Science and Nanotechnology, University of Oslo, P. O. Box 1048 Blindern, NO-0316 Oslo, Norway}

\author{Martin Schaden}
\affiliation{Department of Physics, Rutgers, The State University of New Jersey, Newark, New Jersey 07102, USA}

\author{Iver Brevik}
\affiliation{Department of Energy and Process Engineering, Norwegian University of Science and Technology, NO-7491 Trondheim, Norway}

\author{Drew F. Parsons}
\affiliation{School of Engineering and IT, Murdoch University, 90 South St, Murdoch, WA 6150, Australia}

\author{Kimball A. Milton}
\affiliation{Homer L. Dodge Department of Physics and Astronomy, University of Oklahoma, Norman, Oklahoma 73019, USA}

\author{Oleksandr I. Malyi} 
\email{oleksandr.malyi@smn.uio.no}
\affiliation{Centre for Materials Science and Nanotechnology, University of Oslo, P. O. Box 1048 Blindern, NO-0316 Oslo, Norway}

\author{Mathias Bostr{\"o}m}
\email{Mathias.Bostrom@smn.uio.no}
\affiliation{Centre for Materials Science and Nanotechnology, University of Oslo, P. O. Box 1048 Blindern, NO-0316 Oslo, Norway}

\begin{abstract}
In order to understand why carbon dioxide (CO$_2$) and methane (CH$_4$) molecules interact differently with surfaces, we investigate the Casimir-Polder energy of a linearly polarizable CO$_2$ molecule and an isotropically polarizable CH$_4$ molecule in front of an atomically thin gold film and an amorphous silica slab. We quantitatively analyze how the anisotropy in the polarizability of the molecule influences the van der Waals contribution to the binding energy of the molecule.
\end{abstract}


\maketitle
\section{Introduction}
Underground geological structures, e.g. rocks, shales and depleted coal beds, may contain large fractions of the world's future fossil energy in the form of light natural gas like methane~\cite{Oldenburg}. New production methods like hydraulic fracturing or fracking by CO$_2$ injection allow trapped hydrocarbons to be produced directly from tight source rocks such as shale gas and shale oil systems~\cite{IntechNaturalGas}. The reason that this method works may be partly due to the difference in electron affinities of the CO$_2$ and CH$_4$ molecules~\cite{Babarao}. Much basic research focuses on molecular physisorption/chemisorption and meso-scale transport processes in nanostructured shales~\cite{Abdoulghafour}. By understanding the underlying physical processes, the ultimate aim of such research is to realize a controlled displacement of methane (CH$_4$) by carbon dioxide (CO$_2$) in hydrofractured shale and other formations. The added benefit of such a process is the simultaneous sequestration of the CO$_2$ gas. CO$_2$ gas can be trapped from huge point sources like power plants before it gets released into the atmosphere, and can be conveniently processed and utilized for fracking purposes~\cite{Haszeldine}. 

%------------------------------
\begin{figure}
\includegraphics[width=6cm]{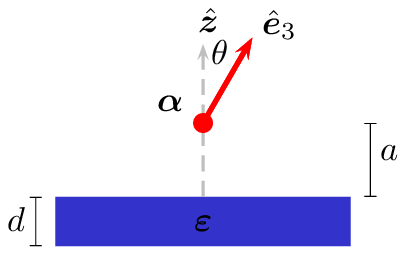}
\includegraphics[width=2cm]{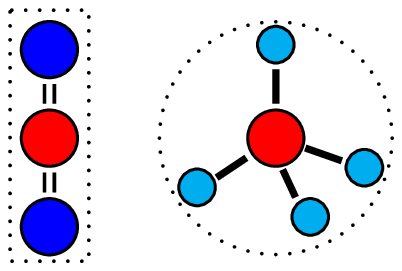}
\caption{(Color online) (Left) Schematic figure of an anisotropically polarizable molecule above a dielectric slab. (Right) Schematic figure showing anisotropy of CO$_2$ molecule and the isotropy of CH$_4$ molecule in their polarizabilities.}
\label{atom-slab}
\end{figure}
%------------------------------

%-------------------------------------------------------------------
From previous studies~\cite{ptPRE, ptCOLSUA} which do not take anisotropy into account, we know that a CO$_2$ molecule and a CH$_4$ molecule adhere to surfaces with very similar van der Waals energies while observed geological phenomena indicate strong preference of CO$_2$ molecule over CH$_4$ molecule in surface adsorption.
In the present work, we explore the physical aspect of the underlying mechanism in order to understand the difference in interaction behavior of CO$_2$ and CH$_4$ molecules with surfaces with the focus being anisotropy in the electric polarizabilities of the respective molecules (see FIG.~\ref{atom-slab}). We address this by studying the interaction energy of an anisotropically polarizable molecule in front of a dielectric slab of finite thickness which is anisotropic only in the direction perpendicular to the slab~\cite{Parashar:2012it, ShajeshPRA2012}. We study the contribution from anisotropic polarizabilities to the Casimir-Polder interaction energy of a CO$_2$ and a CH$_4$ molecule in front of atomically thin gold films and an amorphous silica slab. We choose amorphous silica in particular because surfaces generated by hydraulic fracturing are mostly mineral surfaces like amorphous silica. Amorphous silica is isotropic in nature. We explore the data of the dielectric function of gold available in Ref.~\cite{BoEPJB} to study the variation of interaction energy with film thickness incorporating the effects of anisotropy of the gold film.

In the section II, we present the formalism of the Casimir-Polder interaction energy between a completely anisotropic molecule and a dielectric slab which is anisotropic in the direction perpendicular to the surface. In section III, we briefly summarize the method used for the calculation of the dielectric properties of the slabs. We also briefly describe the procedure used to obtain the anisotropic polarizabilities of the molecules. The dielectric functions of amorphous silica and gold are based on the density functional theory, and the anisotropic polarizabilities of CO$_2$ and CH$_4$  molecules are also obtained from ab initio calculations~\cite{ParsonsA,ParsonsB}. Together the dielectric properties thus obtained are used to determine how the difference in the nature of polarizabilities of CO$_2$ and CH$_4$ distinguish their interaction energies. We present our results in section IV, and end with a few conclusions in section V.

%%%%%%%%%%%%%%%%%%%%%%%%%%%%%%%%%%%%%%%%%%%%%%%%%%%%%%%%%%%%%%%%%%%%%%%%%%%%%%%%%%%%%%%
\section{Formalism}
Consider an anisotropically polarizable molecule described by a frequency dependent molecular polarizability
\begin{equation}
{\bm\alpha}(\omega) = \alpha_1(\omega) \hat{\bf e}_1 \hat{\bf e}_1
+ \alpha_2(\omega) \hat{\bf e}_2 \hat{\bf e}_2
+ \alpha_3(\omega) \hat{\bf e}_3 \hat{\bf e}_3,
\end{equation}
at a distance $a$ above an anisotropically polarizable dielectric slab of thickness $d$ described 
by dielectric permittivity
\begin{equation}
{\bm\varepsilon}(\omega) = 
\varepsilon^\perp(\omega) {\bm 1}_\perp +
\varepsilon^{||}(\omega)\hat{\bf z}\hat{\bf z},
\end{equation}
where $\perp$-components are in the $x$-$y$ plane containing the dielectric slab 
and ${||}$-component is normal to the surface of the slab. See FIG.~\ref{atom-slab}. For an isotropic material like amorphous silica, we can set ${\varepsilon^{\perp}}={\varepsilon^{||}}$. Magnetic permeabilities for both the molecule and the dielectric slab are set to 1.

%-------------------------------------
Here, the principal axes of the molecule are
\begin{subequations}
\begin{eqnarray}
\hat{\bf e}_1 &=& \cos\beta\,\hat{\bm\theta}  
+ \sin\beta \,\hat{\bm\phi}, \\
\hat{\bf e}_2 &=& -\sin\beta\,\hat{\bm\theta}  
+ \cos\beta \,\hat{\bm\phi}, \\
\hat{\bf e}_3 &=& \hat{\bf r}, 
\end{eqnarray}
\end{subequations}
where $\beta$ is the rotation about the unit vector ${\hat{\bm e}_3}$, and $\hat{\bf r}$, $\hat{\bm \theta}$, and $\hat{\bm\phi}$
are the unit vectors in the spherical polar coordinates,
\begin{subequations}
\begin{eqnarray}
\hat{\bf r} &=& \sin\theta \cos\phi \,\hat{\bf x}
+ \sin\theta \sin\phi \,\hat{\bf y} + \cos\theta \,\hat{\bf z}, \\
\hat{\bm \theta} &=& \cos\theta \cos\phi \,\hat{\bf x}
+ \cos\theta \sin\phi \,\hat{\bf y} - \sin\theta \,\hat{\bf z}, \\
\hat{\bm\phi} &=& -\sin\phi \,\hat{\bf x}          
+ \cos\phi \,\hat{\bf y}.
\end{eqnarray}
\end{subequations}

%----------------------------------------------------------------------
Our configuration with an anisotropic molecule above a dielectric slab with isotropic polarizability in the $x$-$y$ plane renders the interaction energy independent of $\phi$. Thus, if necessary, one can choose $\phi=0$, but we will not bother to do so. The Casimir-Polder interaction energy, in this case, at zero temperature in the Fourier transformed space is, \begin{equation}
E =- \hbar c  \int\displaylimits_{-\infty}^\infty \frac{d\zeta}{c}
\int\displaylimits_0^\infty \frac{k_\perp dk_\perp}{2\pi} \int\displaylimits_0^{2\pi}\frac{d\phi_k}{2\pi}
\,\frac{e^{-2\kappa a}}{2\kappa} 
I(i \zeta),
\label{CP12-tp}%
\end{equation} 
which is a generalization of the result given in Ref.~\cite{Parashar:2012it}. The details of the derivation leading to Eq.~(\ref{CP12-tp}) has been omitted for brevity. Here,
\begin{eqnarray}
I(i \zeta)&=&
r^H [\kappa^2 (\hat{\bf k}_\perp\cdot {\bm \alpha} \cdot \hat{\bf k}_\perp) 
                        + k_\perp^2(\hat{\bf z}\cdot{\bm \alpha}\cdot \hat{\bf z})]\nonumber\\
 &&\hspace{0.4cm} -r^E\zeta^2 [(\hat{\bf z}\times\hat{\bf k}_\perp)\cdot{\bm \alpha}\cdot (\hat{\bf z}\times\hat{\bf k}_\perp)].
\label{izeta}
\end{eqnarray} 

The particular choice of $(\hat{\bf k}_\perp,\hat{\bf z}\times\hat{\bf k}_\perp,\hat{\bf z})$ 
basis facilitates separation of TM and TE modes. Specifically,
\begin{subequations}
\begin{eqnarray}
\hat{\bf k}_\perp &=& \cos\phi_k \hat{\bf x}+\sin\phi_k \hat{\bf y}, \\
\hat{\bf z}\times\hat{\bf k}_\perp &=& -\sin\phi_k \hat{\bf x}+\cos\phi_k \hat{\bf y}.
\end{eqnarray}
\end{subequations}
$r^H$ and $r^E$ are the reflection coefficients for TM- and TE-modes:
\begin{subequations}%
\begin{eqnarray}
r^H 
&=& -\left(\frac{\bar{\kappa}^H -\kappa}{\bar{\kappa}^H+\kappa} \right)
\frac{(1- e^{-2\kappa^Hd})} {\left[ 1- 
\left(\frac{\bar{\kappa}^H -\kappa}{\bar{\kappa}^H+\kappa} \right)^2
e^{-2\kappa^Hd}\right]}, 
\\
r^E 
&=& -\left(\frac{\kappa^E -\kappa}{\kappa^E+\kappa} \right)
\frac{(1- e^{-2\kappa^Ed})} {\left[ 1- 
\left(\frac{\kappa^E -\kappa}{\kappa^E+\kappa} \right)^2
e^{-2\kappa^Ed}\right]}, 
\end{eqnarray}%
\label{thick-rs}%
\end{subequations}
where
%---------------------------------------
\begin{subequations}
\begin{alignat}{2}
\kappa^H
&= \sqrt{k_\perp^2 \frac{\varepsilon^\perp}{\varepsilon^{||}}
+ \frac{\zeta^2}{c^2} \varepsilon^\perp }, &\quad
\bar{\kappa}^H &=\frac{\kappa^H}{\varepsilon^\perp}, \\
\kappa^E
&= \sqrt{k_\perp^2 + \frac{\zeta^2}{c^2} \varepsilon^\perp}, &\quad
\kappa&= \sqrt{ k_\perp^2 + \frac{\zeta^2}{c^2}},
\end{alignat}
\end{subequations}
%---------------------------------------
and, $d$ is the thickness of the dielectric slab. The interaction energy after performing the $\phi_k$ 
integration is
\begin{widetext}
\begin{eqnarray}
E &=&
- \hbar c \int\displaylimits_{-\infty}^\infty \frac{d\zeta}{c}
         \int\displaylimits_0^\infty \frac{k_\perp dk_\perp}{2\pi}
         \frac{e^{-2\kappa a}}{2\kappa}
         \bigg\{k_\perp^2 r^H\alpha_3 
               +\left(\frac{\zeta^2}{c^2}(r^H-r^E)+k_\perp^2 r^H\right)
                 \left(\frac{\alpha_1+\alpha_2}{2}\right)\nonumber\\
&&\hspace{2cm}               +\frac12\left(\frac{\zeta^2}{c^2}(r^H-r^E)-k_\perp^2 r^H\right)
                             \left(\alpha_3-\frac{\alpha_1+\alpha_2}{2}
                                 +\left(\frac{\alpha_2-\alpha_1}{2}\right)\cos2\beta\right)\sin^2\theta\bigg\},
\label{Ean-full}
\end{eqnarray}
\end{widetext}
where we have suppressed the frequency dependence. The orientation dependence appears only in the last term, which vanishes for $\theta=0,\pi$.
%----------------------------------------------------------------------
\subsection{Perfect conductor limit}
In the perfect conductor static limit, when $r^H\to1$ and $r^E\to-1$ 
and the molecular polarizability gets a contribution only from the zero frequency, we reproduce 
the known result for the Casimir-Polder energy between an anisotropic molecular and a perfectly conducting slab:
\begin{equation}
E=-\hbar c\frac{\alpha_1+\alpha_2+\alpha_3}{8\pi a^4},
\end{equation}
where last term in Eq.~(\ref{Ean-full}) uniformly integrates to zero. Thus in this case the orientation of 
the molecule has no effect on the interaction energy.

%------------------
\begin{figure}
\includegraphics[width=9cm]{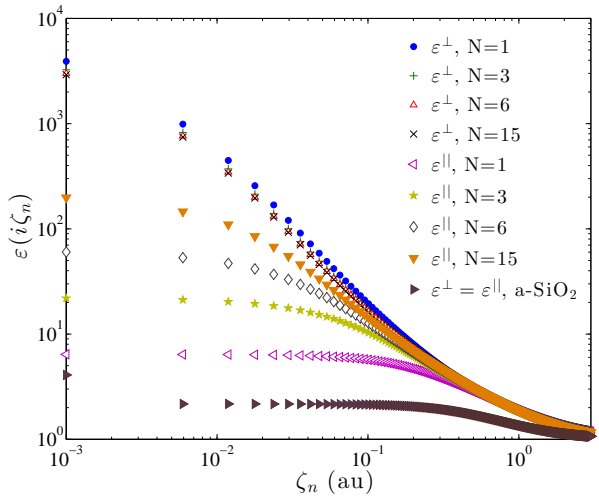}
\caption{(Color online) The perpendicular and parallel dielectric constants for N=1, 3, 6 and 15 atomic layers of gold, and for an amorphous silica slab (written as a-SiO$_2$ in the figure) in terms of the Matsubara frequencies. The perpendicular components of the 3, 6 and 15 atomic-layer thick gold films almost overlap. The dielectric constants at zero frequency are shown on the $y$-axis. $1$ au=$6.57968\times10^{15}$ Hz=$27.212$ eV.}
\label{figdiel}
\end{figure}
%--------------------------------------------------------------------------
\subsection{Temperature dependence}
To account for the temperature ($T$) dependence, we simply replace the integration 
over imaginary frequencies by a summation over discrete Matsubara frequencies ${\zeta_n}$~\cite{Lifshitz,Mahanty2},
\begin{equation}
E =- 2 {k_B}T\sum\limits_{n = 0}^\infty  {'} 
\int_0^\infty dk_\perp k_\perp \,\frac{e^{-2\kappa a}}{2\kappa}  I(i \zeta_n),
\label{CP12-tp_FreeEn}
\end{equation} 
where ${\zeta_n}=2{\pi}{k_B}Tn/{\hbar}$, $k_B$ is the Boltzmann constant, and the prime indicates that the $n=0$ term should be divided by 2. $I(i \zeta_n)$ is given by Eq.~(\ref{izeta}), with ${\zeta}$ replaced by ${\zeta_n}$.

%---------------------------------------------------------------------------------
\subsection{Non-retarded limit}
In the non-retarded limit ${\zeta_n}=0$, the finite temperature interaction energy between an anisotropic molecule at a distance $a$ above an isotropic half-space Eq.~(\ref{Ean-full}) turns out to be,
\begin{widetext}
\begin{eqnarray}
E^{NR}=-2 {k_B}T\sum\limits_{n = 0}^\infty  {'}
         \int\displaylimits_0^\infty dk_\perp k_\perp^2\Delta
         e^{-2k_\perp a}
         \bigg\{\alpha_3+
                 \left(\frac{\alpha_1+\alpha_2}{2}\right)
                                -\frac12
                                 \left(\alpha_3-\frac{\alpha_1+\alpha_2}{2}
                                +\left(\frac{\alpha_2-\alpha_1}{2}\right)\cos2\beta\right)\sin^2\theta\bigg\},
\label{Ean-NR}                                
\end{eqnarray}
\end{widetext}
using $r^H\to\Delta=\frac{\varepsilon-1}{\varepsilon+1}$ and $r^E\to0$. This energy is proportional to $\frac{1}{a^3}$.

%---------------------------------------------------------------------

\begin{figure}
\includegraphics[width=9cm]{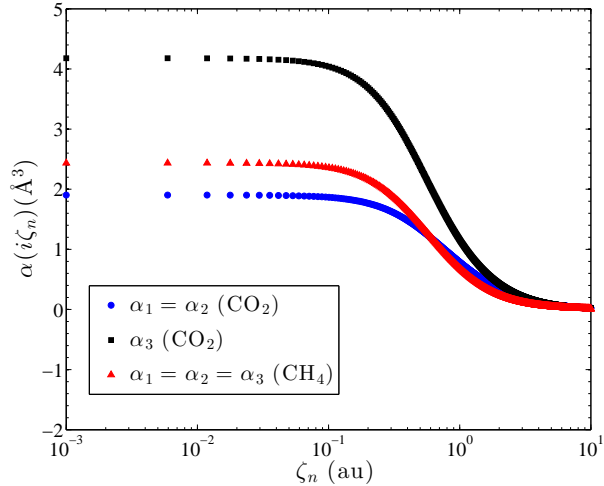}
\caption{(Color online) The anisotropic polarizabilities of CO$_2$ and CH$_4$ in units of {\AA}$^3$ in terms of the Matsubara frequencies. The zero frequency polarizabilities are indicated on the $y$-axis. Note that CO$_2$ is much more polarizable in the transverse direction.}
\label{figpol}
\end{figure}

%--------------------------------------------------------------------------------
\section{Dielectric function and polarizability}
All calculations for amorphous silica were carried out using the Vienna ab initio simulation package (VASP) with the Perdew-Burke-Ernzerhof (PBE)~\cite{S1} functional. Projector augmented wave (PAW) pseudopotentials~\cite{S2,S3} were used to model the effect of core electrons. The non-local parts of the pseudopotentials were treated in the real space for the Born-Oppenheimer molecular dynamics (BOMD) and in the reciprocal space for all other density functional theory (DFT) calculations. The structure of amorphous silica was generated using the BOMD simulations of 72-atom supercell with different annealing-quenching temperature protocols similar to earlier studies~\cite{Sarnthein,Ginhoven}. The dielectric properties of amorphous silica were then calculated using the scissors-operator approximation ($\Delta$=3.6) for PBE calculations. The dielectric function on the imaginary frequency axis was determined using the Kramers-Kronig dispersion relation. The low-energy spectra are verified by calculating the static dielectric constants from the Born effective charges. The static dielectric constant was found to be $4.08\pm0.11$. The details of the 
calculations of anisotropic dielectric functions for gold sheets were presented by 
Bostr{\"o}m et al. in Ref.~\cite{BoEPJB}. We plot the parallel and perpendicular dielectric constants as defined in Eq.~(2) of different thicknesses of gold films, and of amorphous silica  in FIG.~\ref{figdiel}. 

The anisotropic polarizability tensors at imaginary 
frequencies for CO$_2$ and CH$_4$ were calculated using the quantum chemistry package 
MOLPRO~\cite{MOLPRO2008_brief}. Calculations were performed at the coupled clusters, 
singles and doubles (CCSD) level of theory. The correlation-consistent aug-cc-pVQZ 
basis set~\cite{PetersonDunning2002} was used. The geometries of the molecules were 
first optimized by energy minimization before being used in polarizability calculations. All calculations were done at room temperature. We show the anisotropic polarizabilities of CO$_2$ and CH$_4$ molecules in FIG.~\ref{figpol}.

\begin{figure}
\includegraphics[width=9cm]{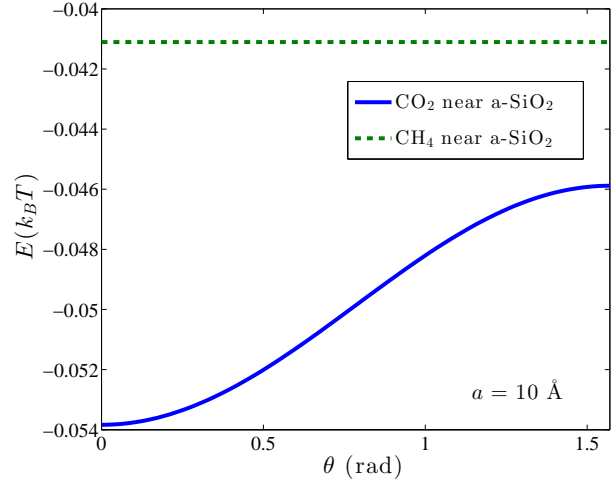}
\caption{(Color online) Comparing the interaction energy of CO$_2$ and CH$_4$ molecules at a distance of $10$ {\AA}  from an amorphous silica slab with respect to ${\theta}$-orientation. We refer to the configuration ${\theta}=0$ as the parallel orientation and ${\theta}$=${\pi}$/2 as the perpendicular orientation. All energies are in units of $k_BT$.}
\label{figu3}
\end{figure}

\begin{figure}
\includegraphics[width=9cm]{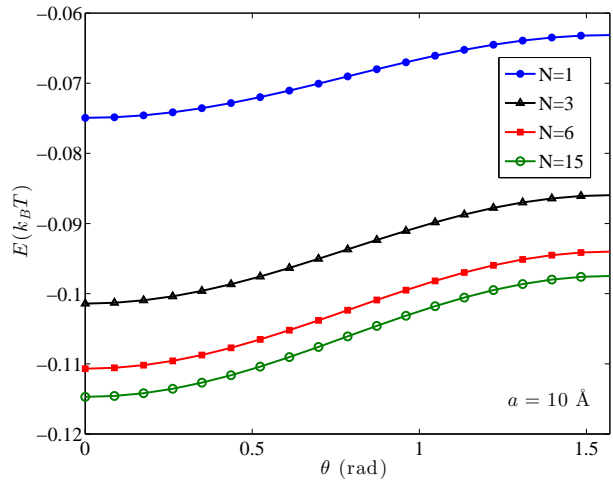}
\caption{(Color online) Interaction energy curves of a CO$_2$ molecule for different ${\theta}$-orientations near gold films of N-atomic-layer thickness. The corresponding energies for a CH$_4$ molecule are -57.9,  -77.8, -84.8 and -87.8 in units of 10$^{-3}$  $k_BT$ for N=1, 3, 6 and 15 respectively.}
\label{figu7}
\end{figure}

\begin{figure}
\includegraphics[width=9cm]{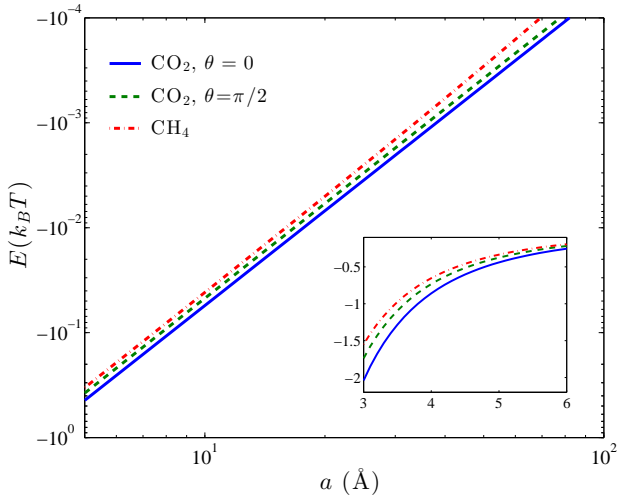}
\caption{(Color online) Comparing the interaction energy of CH$_4$ and CO$_2$ molecules at the parallel (${\theta}=0$) and the perpendicular (${\theta}$=${\pi}$/2) orientations at a varying distance from an amorphous silica slab. The interaction energy for a methane molecule is independent of ${\theta}$-orientation. The inset figure shows the small distance limit. The axis labels and the legends of the outer figure hold for the inset figure as well.}
\label{figu6}
\end{figure}

\begin{figure}
\includegraphics[width=9cm]{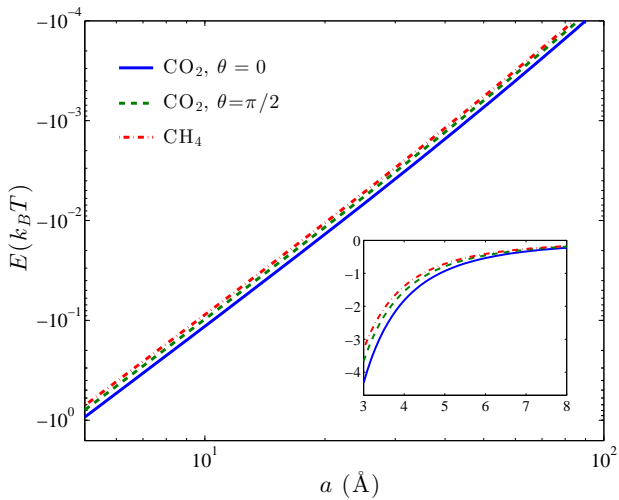}
\caption{(Color online) Comparing the interaction energy of CH$_4$ and CO$_2$ molecules at the parallel (${\theta}=0$) and the perpendicular (${\theta}$=${\pi}$/2) orientations at a varying distance from a 15-atomic-layer thick gold film. }
\label{figu5}
\end{figure}

\section{Numerical results}
For a linear molecule like CO$_2$, the two most notable configurations are the parallel and perpendicular orientations with respect to the dielectric slab, and we are interested in the change in interaction energies in going from one orientation to the other. By parallel orientation, we refer to the configuration in which ${\hat{\bf e}}_3$ is aligned along ${\hat{\bf z}}$ while perpendicular orientation refers to the case when they make an angle $\theta$=$\pi$/2. A CO$_2$ molecule has anisotropy in one direction in its diagonal basis as shown in FIG.~\ref{figpol}. Thus, choosing ${\hat{\bf e}}_3$ along the unique linear axis of the molecule, it is obvious from Eq.~(\ref{Ean-full}) that there is no ${\beta}$ dependent term in the interaction energy. In this particular choice of axes, the unique linear axis of the CO$_2$ molecule is perpendicular to the surface when ${\theta}=0$ (parallel orientation) and parallel to the surface when ${\theta}$=$\pi$/2 (perpendicular orientation). The curves in FIG.~\ref{figu3} show the Casimir-Polder interaction energies of a CO$_2$ and a CH$_4$ molecule for different ${\theta}$-orientations placed at a distance of $10$ {\AA} from an amorphous silica slab. As expected, a methane molecule being highly isotropic shows no change in energy with change in ${\theta}$. A CO$_2$ molecule, on the other hand, exhibits a slight change in the interaction energy at different orientations. The curve in FIG.~\ref{figu3} shows that the CO$_2$ molecule has lower energy at the parallel orientation (${\theta}=0$) than at the perpendicular orientation (${\theta}$=$\pi$/2) near an amorphous silica slab. Thus, the molecule is most stable when its unique linear axis is aligned perpendicular to the slab. Irrespective of their orientations, the magnitude of the energy is larger for a CO$_2$ molecule than for a CH$_4$ molecule. This may play a role in the observed preferential adsorption of CO$_2$ over CH$_4$ molecules on surfaces. These interaction energies are, however, very small compared to $k_BT$. They become comparable when the molecule is very near the surface (see insets of FIGs.~\ref{figu6} and \ref{figu5}). It should be noted that the dielectric continuum picture of our model breaks down at the small distance limit (roughly below 100{\AA}). A more rigorous quantum chemistry calculation will be required to take into account the effects due to surface, bonding, etc.

To make an estimate within our model, we calculate the energy of a system consisting of an amorphous silica slab with CO$_2$ molecules in the parallel orientation at, say, $8$ {\AA} and CH$_4$ molecules at, say, $5$ {\AA} from the slab mimicking the condition when the CO$_2$ gas is being injected. We then consider the reverse system when the CO$_2$ molecules are at $5$ {\AA} and CH$_4$ molecules at $8$ {\AA}. The difference in the interaction energies between the two configurations is $0.078 k_BT$, which is roughly $18$ {\%} compared to the energy in the first configuration. Thus, the second system with CO$_2$ near the surface is more favorable. As stated earlier, at such small separation distances there would be considerable contributions to the interaction energy from other effects. 

In FIG.~\ref{figu7}, we plot curves for the variation of molecule-surface interaction energy with respect to $\theta$ for a CO$_2$ molecule near gold films of different thicknesses. As can be observed from the figure, thicker films give larger magnitudes of interaction energies. Only the interaction energy with the 1-atomic-layer thick gold film displays appreciable difference in comparison with the energy curve for N=15 atomic-layer thick gold film while the interaction energies with the 3 and 6-atomic-layer thick gold films gradually approach that of 15-atomic-layer thick gold film. The energy corresponding to N=15 atomic layers of gold is largest in magnitude. From FIGs.~\ref{figu3} and~\ref{figu7}, we can see that the trends of the energy curves are alike but the molecules have energies larger in magnitude for the more dielectric gold film compared to $100$ {\AA} thick amorphous silica.  We also provide the corresponding energies for a CH$_4$ molecule near gold films of varying thickness in the caption of FIG.~\ref{figu7}. They are $\theta$ independent. 

In FIG.~\ref{figu6}, we fix the CO$_2$ molecule in the parallel and perpendicular orientations and plot the interaction energy with respect to separation distance from the amorphous silica slab on a logarithmic scale. The energy curve for CH$_4$, which is orientation-independent, is also shown. A small difference in the energy curves for the parallel and perpendicular orientations is observed for a CO$_2$ molecule. The interaction energy is larger in magnitude for a CO$_2$ molecule than for a CH$_4$ molecule at all separation distances from the slab owing to greater polarizabilities of CO$_2$ molecule. The curves follow the $\frac{1}{a^3}$ dependence of the non-retarded approximation up to a separation distance of few {\r A}ngst{\"o}ms, and gradually deviates. The inset figure shows the interaction energy in the small molecule-slab separation distance limit (on a linear scale). FIG.~\ref{figu5} shows similar curves near a 15-atomic-layer thick gold film.
%----------------------------------------------------------------------------------
\section{Conclusions}
In this work, we presented a generalized expression for the interaction energy between a completely anisotropic molecule and a dielectric slab polarizable in the direction perpendicular to the surface. Applying this to the specific case of a linearly polarizable CO$_2$ molecule and an isotropically polarizable CH$_4$ molecule, we showed that the anisotropy influences the van der Waals energy to a small degree. The parallel orientation $({\theta}=0)$ is more favored in comparison to the $({\theta}$=${\pi}$/2) perpendicular orientation in the case of a CO$_2$ molecule. In subsequent studies, it will be interesting to incorporate the effects of finite-size of the molecule in which one has to carefully consider different radii of the anisotropic molecule in different directions for determination of the interaction energy for different orientations. In the future, we hope that it will prove possible to transcend the limitations of the continuum approximation, to get more reliable estimates of Casimir-Polder energies at very short distances than we can provide here.

\section*{Acknowledgements}
PT gratefully acknowledges support from the European Commission; this publication reflects the views only of the authors, and the Commission cannot be held responsible for any use which may be made of the information contained therein. PT also acknowledges the Olle Eriksson's Foundation, Sweden (Grant: VT-2014-0001) for supporting a fruitful research visit at the Department of Physics, Southern Illinois University (SIU), Carbondale, USA. PT thanks SIU for hospitality. MB, OIM and CP acknowledge support from the Research Council of Norway (Project: 221469). CP acknowledges support from  the Swedish Research Council (Contract No. C0485101). We acknowledge access to HPC resources at NSC through SNIC/SNAC and at USIT through NOTUR. The work of KAM is supported in part by a grant from the Julian Schwinger Foundation.

%\section{References}
%\begin{comment}

%\end{comment}

\end{document}